# Three-color Photoelectric Observations of UU Cas

**M.I.Kumsiashvili, K.B.Chargeishvili**

E. K. Kharadze National Astrophysical Observatory, Georgia;
e-mail: kumsiashvili@genao.org, ketichargeishvili@yahoo.com

Results of three-color photoelectric UBV observations of UU Cas performed at Abastumani Astrophysical Observatory in 1972-1984 are presented.

Keywords: *stars: close binaries- individual: UU Cas.*

After paper [1] appeared in references we felt a desire to publish this work.

In general, photoelectric observations of the binary system UU Cas were done at Abastumani Astrophysical Observatory according to the plan comprising a group of early spectral type objects (XZ Cep, UU Cas, RY Sct, W Sct, $V_{729}$ Cyg). Their spectral and photometric data are often inconsistent; they are characterized with intensive outflow of matter and complex physical processes.

Three-color photoelectric observations of UU Cas were first performed at Abastumani Observatory in 1972-1984. In each color separately 850 individual observations were carried out. To analyses the light curves variation the methods of theoretical light curves synthesis for close binaries were used as early as in 1992 [2]. Information on the star itself as well as on the performed photoelectric observations is given in the paper mentioned. The fact that, in 1972, 1973 observations of UU Cas (9 nights altogether) were done with a self-recorder and beginning from 1975 a photon counter was used, is noteworthy. Accordingly, shifting of observations by about $0^m.2$ and the amplitude variation of $0^m.1$ in amplitude in V color was observed in these years. Altogether more than 100 observational nights are at our disposal.

Up today there is very poor information on spectral and photometric observations and their studies concerning this star in references; but in 2002 the paper by T.Polushina on the analysis of the light variation for the massive close binary UU Cas was published [1]. It turned out that she did four-color (UBVR) photoelectric observations in 1984-1989. She has 190 individual data in each color separately. In the above analysis average (UBV) photoelectric observations contained in paper [2] are used as well.

Having had solved the variation curves by Lavrov’s method, the author constructed deviations of real light variation curves from model ones. Observations of both seasons (1975-1984, Kumsiashvili) and (1984-1989, Polushina) were used. Despite the fact that these observations are close to each other by their epoch, still they are somehow different; although in both cases deviations of real observations from theoretical curves are significant. They much exceed the observational precision and are about $0^m.1$.

Observational deviations of UU Cas from the model curve in 1975-1984 are presented in Figure.

An opinion turns up that the share of gas structures in this system is comparable to that of the system components itself and the model of gas fluxes is much complex than it is used at the curves analysis with both seasons. As it is seen the circumstellar gas is of multi-component structure and consideration of this fact represents a serious problem of today from the standpoint of constructing a real model of the system.

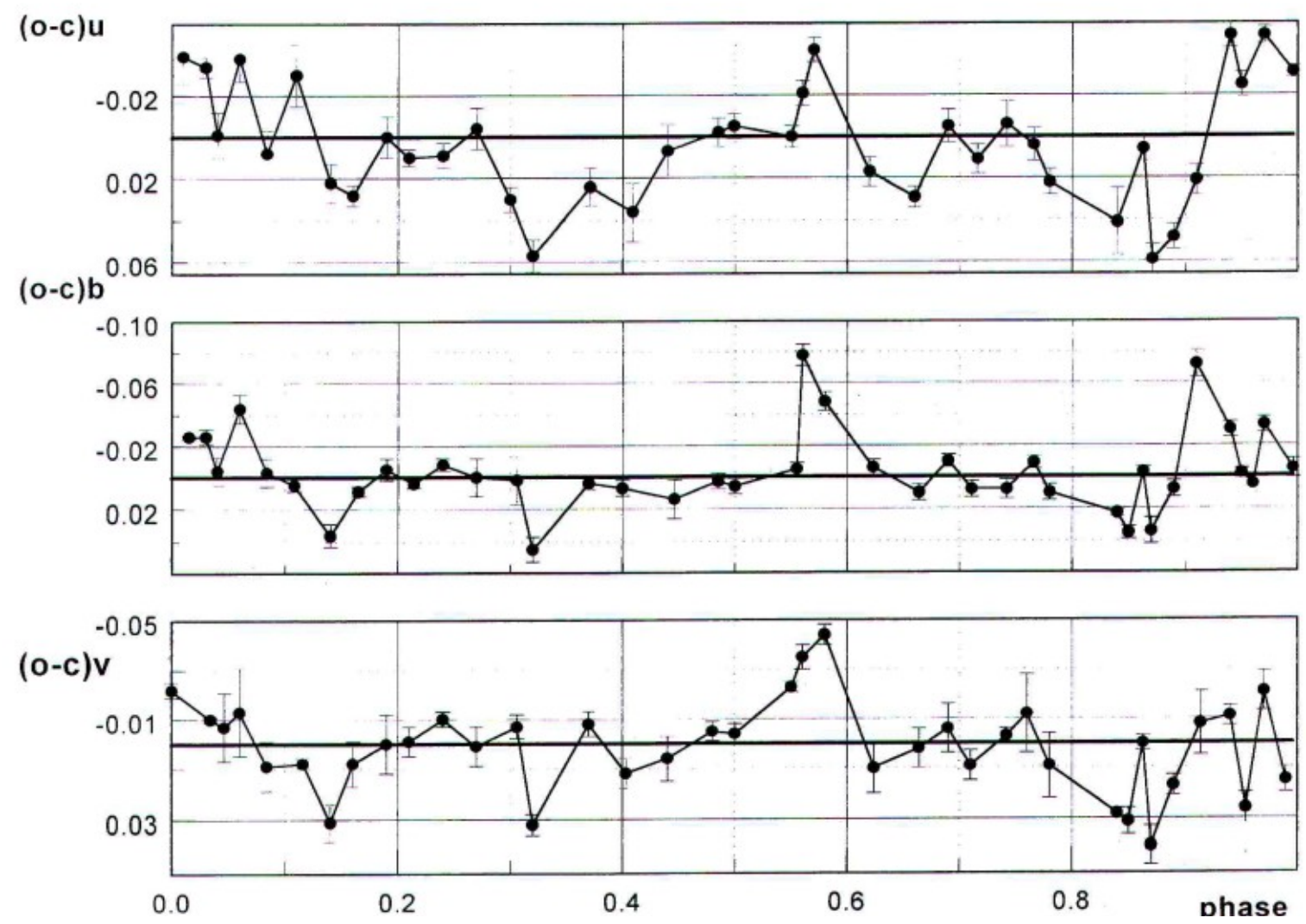

FFig. Deviations of UU Cas observations from the model light curves are displayed in figure.

Consequently, the first thing to do was publishing of our individual photoelectric observations according to nights. The more so that duration of Abastumani observations and abundance of individual data attract one's attention. Accordingly, the light variation from season to season could be discussed and the period, when the action of fluxes is significant, be singled out. Besides, together with photographic observations available, period variability of the system could be discussed. Under conditions of more detailed information on gas fluxes, the light variation curves could be analyzed assuming more complex models and refining our idea about this system. Therefore, aiming at a joint investigation of the above system, we undertook scientificcooperation with the Urals State University as well as with leading specialists of Romania, Poland and Yugoslavia working in our sphere. We hope that this cooperation will be completed with scientific results of interest. Here it should be noted that it is advisable to carry out spectral observations during the total period or, at the worst, at some phases. Necessity for this is also conditioned by the fact that the mass ratio of components is not yet determined in terms of the radial velocity analysis. Therefore estimation of absolute parameters is hindered.

Finally, individual photoelectric observations of UU Cas (BD +60°2629) performwd first at Abastumani observatory using 48 cm AZT-14 telescope are presented. The star BD +60°2631 was used as the comperision one. When drawing light variation curves the phases were calculated by elements [3]:

MinI=JD2428751.72-8.51929.E

The results of observations are given in the Table. The date of observation, moments of observations in Julius days, reduced to the solar center, and phases are listed in columns one, two, three; columns four, five and six show the magnitude differences between the comparison and variable stars in yellow, blue and ultraviolet rays respectively.

| | JD | phase | V | B | U |
|---|---|---|---|---|---|
| 4- 5. XII.72 | 2441656.2195 | .7388 | -.004 | .324 | .560 |
| | .2222 | .7391 | .015 | .329 | .513 |
| | .2250 | .7394 | -.022 | .308 | .547 |
| | .2285 | .7399 | .035 | .312 | .562 |
| | .2327 | .7403 | .002 | .322 | .577 |
| 5- 6. XII.72 | 2441657.1659 | .8499 | .057 | .412 | .648 |
| | .1687 | .8502 | .062 | .450 | .674 |
| | .1749 | .8509 | .088 | .383 | .642 |
| | .1777 | .8513 | .119 | .438 | .645 |
| | .1833 | .8519 | .093 | .437 | .626 |
| | .1860 | .8522 | .069 | .427 | .638 |
| | .1916 | .8529 | .118 | .425 | .666 |
| | .1944 | .8532 | .115 | .412 | .649 |
| | .2020 | .8541 | .100 | .378 | .692 |
| | .2055 | .8545 | .105 | .401 | .618 |
| 7- 8. XII.72 | 2441659.1707 | .852 | .198 | .539 | .753 |
| | .1755 | .858 | .201 | .545 | .751 |
| | .1845 | .868 | .168 | .498 | .759 |
| | .1895 | .874 | .185 | .513 | .757 |
| | .1992 | .886 | .229 | .520 | .757 |
| | .2048 | .892 | .280 | .531 | .741 |
| | .2131 | .902 | .213 | .529 | .773 |
| | .2166 | .906 | .210 | .534 | .796 |
| | .2221 | .912 | .225 | .513 | .832 |
| | .2256 | .917 | .168 | .490 | .758 |
| 9-10. XII.72 | 2441661.1623 | .3190 | .057 | .327 | .573 |
| | .1631 | .3191 | .103 | .334 | .538 |
| | .1700 | .3199 | .049 | .345 | .556 |
| | .1727 | .3202 | .006 | .234 | .285 |
| | .1762 | .3206 | .012 | .326 | .548 |
| | .1797 | .3210 | .010 | .320 | .539 |
| | .1845 | .3216 | .015 | .316 | .547 |
| | .1887 | .3221 | .015 | .314 | .538 |
| | .1929 | .3226 | .002 | .341 | .521 |
| | .1957 | .3229 | .024 | .319 | .572 |
| 28-29. XII.72 | 2441680.2421 | .5586 | .298 | .610 | .823 |
| | .2490 | .5594 | .247 | .581 | .743 |
| | .2553 | .5601 | .283 | .567 | .808 |
| | .2601 | .5607 | .327 | .598 | .805 |
| | .2636 | .5611 | .214 | .644 | .901 |
| 20-21. XI.73 | 2442007.2885 | .9475 | .382 | .669 | .948 |
| | .2927 | .9480 | .403 | .698 | .948 |
| | .2975 | .9486 | .447 | .683 | .963 |
| | .3010 | .9490 | .417 | .691 | 1.033 |
| | .3059 | .9496 | .407 | .675 | .912 |
| | .3100 | .9500 | .416 | .708 | .928 |
| | .3149 | .9506 | .418 | .715 | .944 |
| 26-27. XI.73 | 2442013.3273 | .6564 | .101 | .355 | .621 |
| | .3322 | .6569 | .140 | .415 | .585 |
| | .3377 | .6576 | .082 | .413 | .627 |
| | .3426 | .6581 | .143 | .447 | .610 |
| | .3474 | .6587 | .125 | .483 | .714 |
| | .3516 | .6592 | .104 | .440 | .711 |
| | .3565 | .6598 | .075 | .361 | .643 |
| | .3613 | .6603 | .095 | .389 | .659 |
| | .3655 | .6608 | .112 | .414 | .667 |
| | .3697 | .6613 | .142 | .465 | .719 |
| 27-28. XI.73 | 2442014.2641 | .7663 | .056 | .366 | .548 |
| | .2690 | .7669 | .081 | .365 | .562 |
| | .2738 | .7675 | .068 | .416 | .669 |
| | .2780 | .7679 | .080 | .358 | .573 |
| | .2836 | .7686 | .058 | .357 | .614 |
| | .2877 | .7691 | .055 | .350 | .570 |
| | .2919 | .7696 | .067 | .355 | .617 |
| | .2961 | .7701 | .075 | .330 | .594 |
| | .3009 | .7706 | .084 | .378 | .597 |
| | .3051 | .7711 | .084 | .374 | .616 |
| 16-17. XII.73 | 2442033.1807 | .9868 | .505 | .799 | 1.000 |
| | .1848 | .9872 | .597 | .798 | 1.046 |
| | .1890 | .9877 | .502 | .783 | 1.051 |
| | .1932 | .9882 | .495 | .783 | .993 |
| | .1980 | .9888 | .566 | .817 | .952 |
| | .2057 | .9897 | .507 | .840 | 1.084 |
| | .2105 | .9902 | .578 | .824 | .943 |
| | .2147 | .9907 | .507 | .868 | .953 |
| | .2189 | .9912 | .474 | .769 | .917 |
| | .2258 | .9920 | .511 | .754 | .936 |
| | .2307 | .9926 | .563 | .854 | 1.079 |
| | .2355 | .9932 | .470 | .789 | 1.118 |
| | .2439 | .9942 | .541 | .715 | .822 |
| | .2480 | .9947 | .539 | .699 | .864 |
| | .2515 | .9951 | .615 | .842 | .994 |
| | .2661 | .9968 | .532 | .810 | .948 |
| | .2703 | .9973 | .499 | .815 | 1.157 |
| | .2737 | .9977 | .495 | .780 | 1.054 |
| | .2786 | .9982 | .429 | .753 | 1.070 |
| | .2855 | .9991 | .621 | .827 | 1.091 |
| | .2897 | .9996 | .411 | .706 | 1.088 |
| | .2939 | .0000 | .438 | .788 | 1.127 |
| | .2973 | .0004 | .454 | .795 | .879 |
| | .3022 | .0010 | .628 | .783 | .980 |
| | .3064 | .0015 | .664 | .662 | .644 |
| | .3112 | .0021 | .448 | .499 | .877 |
| | .3154 | .0026 | .616 | .843 | 1.359 |
| 30-31.VIII.75 | 2442655.3094 | .0126 | .256 | .650 | .839 |
| | .3122 | .0130 | .283 | .640 | .782 |
| | .3164 | .0135 | .271 | .632 | .802 |
| | .3192 | .0140 | .280 | .650 | .784 |
| | .3254 | .0145 | .275 | .638 | .834 |
| | .3289 | .0149 | .272 | .646 | .818 |
| | .3324 | .0153 | .261 | .630 | .829 |
| | .3351 | .0157 | .250 | .620 | .813 |
| | .3393 | .0161 | .246 | .647 | .784 |
| | .3421 | .0165 | .264 | .641 | .770 |
| 31- 1. IX.75 | 2442656.3449 | .1342 | .021 | .390 | .571 |
| | .3483 | .1346 | .028 | .396 | .556 |
| | .3525 | .1351 | .031 | .413 | .551 |
| | .3560 | .1355 | .036 | .411 | .523 |
| | .3594 | .1359 | .029 | .418 | .571 |
| | .3629 | .1363 | .024 | .396 | .543 |
| | .3664 | .1367 | .031 | .410 | .567 |
| | .3692 | .1370 | .000 | .394 | .547 |
| | .3726 | .1374 | .000 | .380 | .513 |
| | .3754 | .1378 | .013 | .390 | .520 |
| 7- 8. IX.75 | 2442663.3391 | .9552 | .251 | .625 | .823 |
| | .3419 | .9555 | .248 | .637 | .818 |
| | .3454 | .9559 | .242 | .649 | .808 |
| | .3483 | .9562 | .244 | .632 | .822 |
| | .3516 | .9566 | .248 | .645 | .783 |
| 30-31. X.75 | 2442716.2270 | .1632 | -.042 | .327 | .478 |
| | .2298 | .1635 | -.039 | .329 | .443 |
| | .2333 | .1639 | -.043 | .327 | .495 |
| | .2367 | .1643 | -.046 | .320 | .486 |
| | .2409 | .1648 | -.053 | .322 | .478 |
| | .2437 | .1651 | -.061 | .312 | .451 |
| | .2472 | .1656 | -.059 | .309 | .459 |
| | .2499 | .1659 | -.049 | .316 | .455 |
| | .2534 | .1663 | -.039 | .320 | .440 |
| | .2569 | .1667 | -.049 | .324 | .438 |
| 2- 3. XII.75 | 2442749.2056 | .0342 | .264 | .651 | .820 |
| | .2080 | .0346 | .244 | .635 | .815 |
| | .2146 | .0353 | .253 | .625 | .816 |
| | .2181 | .0357 | .251 | .608 | .801 |
| | .2216 | .0361 | .250 | .643 | .791 |
| | .2250 | .0365 | .259 | .636 | .797 |
| | .2285 | .0369 | .261 | .633 | .814 |
| | .2313 | .0373 | .246 | .642 | .785 |
| | .2354 | .0377 | .262 | .645 | .781 |
| | .2389 | .0381 | .242 | .652 | .812 |
| | .2431 | .0386 | .230 | .634 | .841 |
| | .2459 | .0390 | .261 | .612 | .849 |
| | .2493 | .0394 | .262 | .606 | .820 |
| | .2556 | .0401 | .232 | .631 | .816 |
| | .2591 | .0405 | .228 | .621 | .780 |
| | .2632 | .0410 | .232 | .620 | .870 |
| | .2674 | .0415 | .241 | .613 | .809 |
| | .2709 | .0419 | .222 | .604 | .791 |
| | .2743 | .0423 | .226 | .623 | .798 |
| | .2771 | .0426 | .241 | .639 | .815 |
| | .2834 | .0434 | .222 | .605 | .816 |
| | .2868 | .0438 | .243 | .624 | .778 |
| | .2903 | .0442 | .219 | .618 | .760 |
| | .2931 | .0445 | .230 | .637 | .778 |
| | .2980 | .0451 | .227 | .633 | .784 |
| | .3021 | .0456 | .217 | .628 | .813 |
| | .3056 | .0460 | .238 | .605 | .801 |
| | .3084 | .0463 | .231 | .638 | .805 |
| | .3118 | .0467 | .227 | .618 | .792 |
| | .3153 | .0471 | .232 | .591 | .791 |
| | .3188 | .0475 | .232 | .619 | .795 |
| 5- 6. XII.75 | 2442752.2798 | .3951 | .025 | .384 | .567 |
| | .2826 | .3954 | .023 | .413 | .554 |
| | .2860 | .3958 | .021 | .408 | .546 |
| | .2895 | .3962 | .024 | .412 | .572 |
| | .2930 | .3966 | .017 | .405 | .593 |
| | .2958 | .3970 | .043 | .423 | .577 |
| | .3006 | .3975 | .045 | .418 | .546 |
| | .3034 | .3979 | .034 | .423 | .554 |
| | .3069 | .3983 | .039 | .434 | .567 |
| | .3096 | .3986 | .041 | .424 | .577 |
| | .3138 | .3991 | .039 | .419 | .575 |
| 5- 6.VIII.76 | 2442996.4324 | .0539 | .073 | .451 | .729 |
| | .4352 | .0542 | .073 | .508 | .689 |
| | .4394 | .0547 | .077 | .522 | .646 |

| | | | | | | |
|---|---|---|---|---|---|---|
| | | .4422 | .0550 | .079 | .516 | .685 |
| | | .4456 | .0554 | .081 | .519 | .660 |
| | | .4547 | .0565 | .054 | .481 | .659 |
| | | .4609 | .0572 | .054 | .465 | .636 |
| | | .4644 | .0576 | .055 | .502 | .628 |
| | | .4686 | .0581 | .042 | .492 | .671 |
| | | .4720 | .0585 | .053 | .482 | .688 |
| | | .4748 | .0589 | .075 | .495 | .666 |
| | | .4783 | .0593 | .076 | .491 | .652 |
| | | .4824 | .0598 | .058 | .505 | .650 |
| | | .4859 | .0602 | .048 | .513 | .631 |
| | | .4894 | .0606 | .042 | .484 | .658 |
| 12-13. | XI.76 | 2443095.2282 | .6506 | -.076 | .365 | .531 |
| | | .2317 | .6510 | -.079 | .369 | .548 |
| | | .2386 | .6518 | -.061 | .357 | .533 |
| | | .2435 | .6524 | -.073 | .360 | .557 |
| | | .2476 | .6529 | -.077 | .380 | .546 |
| | | .2511 | .6533 | -.072 | .367 | .505 |
| | | .2553 | .6538 | -.053 | .368 | .525 |
| | | .2588 | .6542 | -.074 | .366 | .529 |
| 13-14. | XI.76 | 2443096.2560 | .7712 | -.152 | .291 | .451 |
| | | .2595 | .7717 | -.151 | .306 | .441 |
| | | .2629 | .7721 | -.154 | .321 | .452 |
| | | .2664 | .7725 | -.156 | .305 | .470 |
| | | .2699 | .7729 | -.154 | .302 | .440 |
| | | .2733 | .7733 | -.153 | .298 | .487 |
| | | .2768 | .7737 | -.151 | .294 | .464 |
| | | .2796 | .7740 | -.217 | .288 | .464 |
| | | .2831 | .7744 | -.206 | .302 | .481 |
| | | .2865 | .7748 | -.153 | .284 | .464 |
| 14-15. | XI.76 | 2443097.2442 | .8872 | .022 | .433 | .648 |
| | | .2476 | .8876 | .007 | .452 | .626 |
| | | .2511 | .8880 | .021 | .455 | .610 |
| | | .2539 | .8884 | .007 | .458 | .626 |
| | | .2574 | .8888 | .010 | .449 | .655 |
| | | .2602 | .8891 | .021 | .456 | .627 |
| | | .2636 | .8895 | .009 | .459 | .640 |
| | | .2671 | .8899 | .009 | .460 | .670 |
| | | .2706 | .8903 | .032 | .454 | .652 |
| | | .2733 | .8907 | .028 | .464 | .650 |
| 15-16. | XI.76 | 2443098.2442 | .0046 | .275 | .706 | .897 |
| | | .2476 | .0050 | .269 | .721 | .902 |
| | | .2511 | .0054 | .274 | .730 | .941 |
| | | .2539 | .0058 | .265 | .716 | .934 |
| | | .2581 | .0063 | .264 | .704 | .923 |
| | | .2609 | .0066 | .274 | .721 | .920 |
| | | .2643 | .0070 | .256 | .716 | .920 |
| | | .2678 | .0074 | .263 | .720 | .925 |
| | | .2713 | .0078 | .259 | .731 | .915 |
| | | .2740 | .0081 | .267 | .724 | .944 |
| | | .2782 | .0086 | .257 | .729 | .886 |
| | | .2810 | .0089 | .270 | .713 | .940 |
| | | .2845 | .0094 | .267 | .711 | .887 |
| | | .2879 | .0098 | .271 | .711 | .902 |
| | | .2914 | .0102 | .271 | .715 | .905 |
| 17-18. | XII.76 | 2443130.2646 | .7632 | -.133 | .334 | .455 |
| | | .2674 | .7635 | -.196 | .301 | .435 |
| | | .2709 | .7639 | -.144 | .319 | .475 |
| | | .2757 | .7645 | -.047 | .346 | .468 |

| | | | | | |
|---|---|---|---|---|---|
| | .2827 | .7653 | -.014 | .330 | .470 |
| | .2861 | .7657 | -.009 | .330 | .466 |
| | .2896 | .7661 | -.008 | .341 | .437 |
| | .2931 | .7665 | -.020 | .333 | .492 |
| | .2965 | .7669 | -.014 | .344 | .496 |
| | .3007 | .7674 | -.013 | .361 | .483 |
| 18-19. XII.76 | 2443131.2327 | .8768 | .068 | .443 | .573 |
| | .2361 | .8772 | .074 | .450 | .562 |
| | .2396 | .8776 | .059 | .439 | .624 |
| | .2431 | .8781 | .074 | .427 | .590 |
| | .2465 | .8785 | .063 | .438 | .589 |
| | .2500 | .8789 | .069 | .436 | .593 |
| | .2535 | .8793 | .076 | .437 | .612 |
| | .2570 | .8797 | .065 | .443 | .572 |
| | .2604 | .8801 | .061 | .444 | .588 |
| | .2632 | .8804 | .071 | .440 | .599 |
| 19-20. XII.76 | 2443132.2396 | .9950 | .316 | .711 | .884 |
| | .2424 | .9954 | .335 | .730 | .870 |
| | .2459 | .9958 | .334 | .711 | .889 |
| | .2486 | .9961 | .337 | .714 | .873 |
| | .2521 | .9965 | .353 | .707 | .886 |
| | .2549 | .9968 | .330 | .709 | .885 |
| | .2591 | .9973 | .327 | .720 | .888 |
| | .2618 | .9976 | .336 | .717 | .870 |
| | .2660 | .9981 | .332 | .728 | .877 |
| | .2688 | .9985 | .354 | .718 | .877 |
| | .2722 | .9989 | .324 | .671 | .883 |
| | .2750 | .9992 | .343 | .695 | .897 |
| | .2785 | .9996 | .359 | .739 | .873 |
| | .2813 | .9999 | .342 | .696 | .841 |
| | .2847 | .0003 | .364 | .723 | .829 |
| 17-18. VII.77 | 2443342.4328 | .6676 | -.055 | .352 | .490 |
| | .4355 | .6680 | -.050 | .343 | .484 |
| | .4398 | .6685 | -.062 | .355 | .485 |
| | .4467 | .6693 | -.038 | .331 | .477 |
| | .4481 | .6694 | -.041 | .356 | .484 |
| | .4523 | .6699 | -.061 | .341 | .502 |
| | .4558 | .6703 | -.051 | .351 | .499 |
| | .4592 | .6707 | -.041 | .345 | .485 |
| | .4620 | .6711 | -.046 | .327 | .487 |
| 13-14. IX.77 | 2443400.4447 | .4771 | .187 | .559 | .687 |
| | .4474 | .4774 | .177 | .536 | .659 |
| | .4507 | .4778 | .171 | .544 | .678 |
| | .4537 | .4782 | .177 | .576 | .708 |
| | .4641 | .4794 | .185 | .561 | .712 |
| | .4669 | .4797 | .165 | .544 | .685 |
| 18-19. IX.77 | 2443405.3963 | .0583 | .182 | .566 | .746 |
| | .3997 | .0587 | .181 | .543 | .742 |
| | .4053 | .0594 | .199 | .570 | .726 |
| | .4088 | .0598 | .190 | .560 | .725 |
| | .4213 | .0613 | .185 | .576 | .741 |
| | .4247 | .0617 | .188 | .585 | .772 |
| | .4289 | .0622 | .168 | .543 | .745 |
| 19-20. IX.77 | 2443406.3449 | .1697 | -.081 | .288 | .466 |
| | .3476 | .1700 | -.080 | .294 | .473 |
| | .3511 | .1704 | -.088 | .299 | .469 |
| | .3553 | .1709 | -.074 | .310 | .441 |
| | .3594 | .1714 | -.082 | .300 | .448 |

| | | | | | | |
|---|---|---|---|---|---|---|
| | | .3629 | .1718 | -.086 | .302 | .454 |
| | | .3678 | .1724 | -.081 | .299 | .446 |
| | | .3713 | .1728 | -.079 | .289 | .447 |
| | | .3747 | .1732 | -.072 | .299 | .439 |
| | | .3782 | .1736 | -.083 | .314 | .435 |
| 20-21. | IX.77 | 2443407.4429 | .2986 | -.116 | .265 | .413 |
| | | .4457 | .2989 | -.109 | .263 | .417 |
| | | .4491 | .2993 | -.102 | .276 | .447 |
| | | .4519 | .2996 | -.121 | .272 | .425 |
| | | .4554 | .3000 | -.118 | .283 | .435 |
| | | .4582 | .3004 | -.104 | .282 | .436 |
| | | .4623 | .3009 | -.119 | .266 | .448 |
| | | .4651 | .3012 | -.132 | .263 | .428 |
| | | .4693 | .3017 | -.127 | .253 | .415 |
| | | .4714 | .3019 | -.118 | .260 | .412 |
| 21-22. | IX.77 | 2443408.5193 | .4249 | .109 | .467 | .832 |
| | | .5227 | .4253 | .107 | .498 | .597 |
| | | .5269 | .4258 | .098 | .473 | .625 |
| | | .5297 | .4261 | .117 | .473 | .630 |
| | | .5332 | .4266 | .112 | .492 | .617 |
| | | .5359 | .4269 | .121 | .485 | .596 |
| | | .5401 | .4274 | .113 | .483 | .626 |
| | | .5429 | .4277 | .110 | .480 | .610 |
| | | .5464 | .4281 | .115 | .475 | .635 |
| | | .5482 | .4283 | .111 | .468 | .616 |
| 22-23. | IX.77 | 2443409.4887 | .5387 | .119 | .530 | .663 |
| | | .4922 | .5391 | .128 | .507 | .670 |
| | | .4957 | .5395 | .129 | .504 | .648 |
| | | .4984 | .5398 | .122 | .509 | .613 |
| | | .5012 | .5402 | .124 | .507 | .639 |
| | | .5040 | .5405 | .123 | .506 | .645 |
| | | .5068 | .5408 | .120 | .487 | .674 |
| | | .5095 | .5412 | .118 | .494 | .630 |
| | | .5151 | .5418 | .105 | .497 | .637 |
| | | .5179 | .5421 | .102 | .495 | .648 |
| | | .5207 | .5425 | .105 | .503 | .656 |
| | | .5434 | .5428 | .109 | .508 | .641 |
| 6- 7. | X.77 | 2443423.3141 | .1615 | -.035 | .325 | .476 |
| | | .3169 | .1619 | -.041 | .321 | .469 |
| | | .3197 | .1622 | -.045 | .335 | .471 |
| | | .3224 | .1625 | -.050 | .337 | .461 |
| | | .3252 | .1628 | -.048 | .318 | .480 |
| | | .3280 | .1632 | -.054 | .326 | .495 |
| | | .3315 | .1636 | -.046 | .326 | .497 |
| | | .3343 | .1639 | -.050 | .332 | .482 |
| | | .3363 | .1642 | -.061 | .338 | .455 |
| | | .3412 | .1647 | -.064 | .334 | .466 |
| 7- 8. | X.77 | 2443424.3156 | .2791 | -.078 | .295 | .437 |
| | | .3198 | .2796 | -.079 | .310 | .462 |
| | | .3225 | .2799 | -.074 | .301 | .441 |
| | | .3274 | .2805 | -.078 | .321 | .442 |
| | | .3302 | .2808 | -.064 | .319 | .451 |
| | | .3337 | .2812 | -.064 | .308 | .444 |
| | | .3364 | .2815 | -.070 | .311 | .434 |
| | | .3385 | .2818 | -.078 | .302 | .455 |
| | | .3420 | .2822 | -.073 | .309 | .460 |
| 8- 9. | X.77 | 2443425.3274 | .3979 | .062 | .426 | .565 |
| | | .3309 | .3983 | .053 | .427 | .568 |
| | | .3337 | .3986 | .060 | .439 | .610 |
| | | .3350 | .3988 | .057 | .444 | .623 |
| | | .3385 | .3992 | .067 | .416 | .579 |
| | | .3413 | .3995 | .080 | .430 | .599 |
| | | .3448 | .3999 | .072 | .447 | .609 |
| | | .3469 | .4002 | .067 | .427 | .627 |
| | | .3489 | .4004 | .075 | .434 | .608 |
| | | .3538 | .4010 | .062 | .455 | .596 |
| 11-12. | X.77 | 2443428.3296 | .7503 | -.106 | .279 | .449 |
| | | .3338 | .7508 | | | |
| | | .3372 | .7512 | -.107 | .295 | .407 |
| | | .3400 | .7515 | -.107 | .295 | .428 |
| | | .3428 | .7518 | -.113 | .278 | .407 |
| | | .3456 | .7521 | -.113 | .299 | .416 |
| | | .3483 | .7525 | -.096 | .298 | .408 |
| | | .3511 | .7528 | -.119 | .288 | .420 |
| | | .3539 | .7531 | -.097 | .281 | .403 |
| | | .3560 | .7534 | -.106 | .261 | .405 |
| 12-13. | X.77 | 2443429.2990 | .8641 | -.015 | .343 | .508 |
| | | .3011 | .8643 | -.024 | .343 | .499 |
| | | .3046 | .8647 | -.022 | .354 | .487 |
| | | .3074 | .8650 | -.026 | .351 | .500 |
| | | .3101 | .8654 | -.026 | .364 | .502 |
| | | .3129 | .8657 | -.033 | .371 | .520 |
| | | .3157 | .8660 | -.030 | .360 | .512 |
| | | .3178 | .8663 | -.024 | .347 | .534 |
| | | .3206 | .8666 | -.013 | .374 | .520 |
| | | .3233 | .8669 | -.022 | .376 | .501 |
| 13-14. | X.77 | 2443430.3164 | .9835 | .237 | .610 | .804 |
| | | .3185 | .9837 | .230 | .619 | .812 |
| | | .3220 | .9841 | .248 | .622 | .796 |
| | | .3247 | .9845 | .254 | .631 | .805 |
| | | .3275 | .9848 | .262 | .639 | .813 |
| | | .3296 | .9850 | .256 | .623 | .808 |
| | | .3338 | .9855 | .261 | .643 | .814 |
| | | .3372 | .9859 | .249 | .636 | .824 |
| | | .3393 | .9862 | .253 | .639 | .809 |
| | | .3421 | .9865 | .257 | .638 | .807 |
| | | .3449 | .9868 | .262 | .639 | .780 |
| 19-20. | X.77 | 2443436.4068 | .6984 | -.149 | .235 | .409 |
| | | .4096 | .6987 | -.120 | .251 | .403 |
| | | .4123 | .6990 | -.121 | .247 | .409 |
| | | .4151 | .6993 | -.133 | .250 | .385 |
| | | .4186 | .6998 | -.144 | .256 | .392 |
| | | .4214 | .7001 | -.145 | .266 | .409 |
| | | .4248 | .7005 | -.150 | .248 | .401 |
| | | .4276 | .7008 | -.142 | .270 | .393 |
| | | .4304 | .7011 | -.147 | .258 | .356 |
| | | .4332 | .7015 | -.132 | .261 | .386 |
| 1- 2. | XI.77 | 2443449.2923 | .2031 | -.094 | .255 | .438 |
| | | .2951 | .2035 | -.096 | .262 | .419 |
| | | .2978 | .2038 | -.090 | .291 | .442 |
| | | .3006 | .2041 | -.101 | .290 | .442 |
| | | .3041 | .2045 | -.097 | .269 | .423 |
| | | .3069 | .2049 | -.086 | .268 | .394 |
| | | .3097 | .2052 | -.083 | .274 | .415 |
| | | .3124 | .2055 | -.099 | .271 | .440 |
| | | .3153 | .2058 | -.109 | .285 | .404 |
| | | .3180 | .2062 | -.093 | .286 | .388 |

| | | | | | |
|---|---|---|---|---|---|
| 12-13. I.77 | 2443521.2980 | .6630 | -.073 | .283 | .446 |
| | .3015 | .6634 | -.068 | .314 | .448 |
| | .3050 | .6638 | -.062 | .307 | .447 |
| | .3077 | .6641 | -.049 | .333 | .454 |
| | .3133 | .6648 | -.057 | .316 | .454 |
| | .3161 | .6651 | -.068 | .334 | .443 |
| | .3196 | .6655 | -.070 | .335 | .445 |
| | .3230 | .6659 | -.045 | .258 | .459 |
| | .3258 | .6662 | -.325 | .325 | .451 |
| | .3293 | .6666 | -.060 | .324 | .456 |
| 14-15. I.77 | 2443523.4188 | .9119 | .077 | .400 | .478 |
| | .4222 | .9123 | .086 | .368 | .496 |
| | .4264 | .9128 | .043 | .395 | .455 |
| | .4292 | .9131 | .061 | .415 | .466 |
| | .4340 | .9137 | .060 | .452 | .488 |
| | .4368 | .9140 | .054 | .414 | .454 |
| | .4403 | .9144 | .066 | .431 | .393 |
| | .4431 | .9148 | .091 | .418 | .447 |
| | .4458 | .9151 | .086 | .425 | .449 |
| | .4486 | .9154 | .100 | .420 | .427 |
| 30-31. VII.78 | 2443720.4063 | .0344 | .220 | .608 | .824 |
| | .4111 | .0350 | .216 | .648 | .781 |
| | .4146 | .0354 | .232 | .620 | .750 |
| | .4174 | .0357 | .241 | .608 | .732 |
| | .4208 | .0361 | .237 | .615 | .753 |
| | .4243 | .0365 | .237 | .597 | .783 |
| | .4278 | .0370 | .232 | .617 | .832 |
| | .4306 | .0373 | .235 | .588 | .817 |
| | .4347 | .0378 | .233 | .627 | .784 |
| | .4375 | .0381 | .236 | .627 | .773 |
| | .4410 | .0385 | .226 | .602 | .745 |
| | .4444 | .0389 | .222 | .607 | .766 |
| | .4473 | .0393 | .245 | .629 | .742 |
| 9-10.VIII.78 | 2443730.4603 | .2146 | -.100 | .293 | .423 |
| | .4631 | .2149 | -.090 | .289 | .404 |
| | .4659 | .2152 | -.080 | .278 | .398 |
| | .4687 | .2156 | -.096 | .279 | .390 |
| | .4714 | .2159 | -.102 | .306 | .439 |
| 10-11.VIII.78 | 2443731.4653 | .3325 | -.061 | .314 | .433 |
| | .4681 | .3329 | -.055 | .353 | .429 |
| | .4708 | .3332 | -.063 | .332 | .511 |
| | .4736 | .3335 | -.059 | .327 | .430 |
| | .4764 | .3338 | -.041 | .302 | .513 |
| 9-10. IX.78 | 2443761.4361 | .8505 | -.004 | .370 | .505 |
| | .4389 | .8509 | -.031 | .353 | .527 |
| | .4417 | .8512 | -.039 | .357 | .541 |
| 10-11. IX.78 | 2443762.4328 | .9675 | .247 | .621 | .791 |
| | .4355 | .9678 | .254 | .637 | .848 |
| | .4383 | .9682 | .245 | .636 | .828 |
| | .4411 | .9685 | .254 | .680 | .843 |
| | .4439 | .9688 | .271 | .648 | .858 |
| | .4466 | .9691 | .272 | .665 | .863 |
| | .4501 | .9696 | .259 | .656 | .803 |
| | .4529 | .9699 | .273 | .659 | .818 |
| | .4564 | .9703 | .240 | .646 | .818 |
| | .4599 | .9707 | .254 | .640 | .831 |
| | .4626 | .9710 | .256 | .664 | .789 |

| | | | | | | |
|---|---|---|---|---|---|---|
| | | .4654 | .9714 | .268 | .649 | .807 |
| | | .4689 | .9718 | .252 | .665 | .776 |
| | | .4716 | .9721 | .254 | .653 | .786 |
| | | .4744 | .9724 | .258 | .635 | .825 |
| 11-12. | IX.78 | 2443763.4758 | .0900 | .060 | .448 | .570 |
| | | .4786 | .0903 | .076 | .495 | .600 |
| | | .4814 | .0906 | .084 | .458 | .678 |
| | | .4841 | .0909 | .066 | .434 | .665 |
| | | .4869 | .0913 | .086 | .472 | .631 |
| | | .4897 | .0916 | .066 | .438 | .614 |
| | | .4925 | .0919 | .068 | .457 | .635 |
| | | .4953 | .0922 | .057 | .429 | .640 |
| | | .4980 | .0926 | .081 | .469 | .625 |
| | | .5015 | .0930 | .064 | .460 | .615 |
| 12-13. | IX.78 | 2443764.4947 | .2096 | -.127 | .291 | .437 |
| | | .4974 | .2099 | -.119 | .293 | .440 |
| | | .5002 | .2102 | -.100 | .299 | .413 |
| | | .5030 | .2105 | -.103 | .301 | .404 |
| | | .5058 | .2109 | -.117 | .300 | .420 |
| | | .5086 | .2112 | -.108 | .294 | .432 |
| | | .5113 | .2115 | -.108 | .257 | .432 |
| 13-14. | IX.78 | 2443765.4863 | .3260 | -.609 | .120 | .533 |
| | | .4898 | .3264 | -.034 | .369 | .725 |
| | | .4961 | .3271 | -.014 | .378 | .494 |
| | | .4995 | .3275 | -.030 | .362 | .503 |
| | | .5023 | .3278 | -.032 | .374 | .496 |
| | | .5051 | .3282 | -.011 | .359 | .520 |
| | | .5079 | .3285 | -.037 | .374 | .496 |
| | | .5127 | .3291 | -.037 | .347 | .460 |
| | | .5148 | .3293 | -.043 | .348 | .487 |
| 26-27. | IX.78 | 2443778.3736 | .8387 | -.029 | .356 | .446 |
| | | .3764 | .8390 | -.021 | .351 | .461 |
| | | .3792 | .8393 | -.028 | .372 | .550 |
| | | .3820 | .8397 | -.028 | .384 | .564 |
| | | .3847 | .8400 | -.031 | .364 | .514 |
| | | .3875 | .8403 | -.022 | .339 | .497 |
| | | .3903 | .8406 | -.028 | .334 | .531 |
| 27-28. | IX.78 | 2443779.3445 | .9526 | .243 | .633 | .801 |
| | | .3472 | .9529 | .244 | .643 | .805 |
| | | .3507 | .9534 | .241 | .622 | .764 |
| | | .3535 | .9537 | .235 | .623 | .762 |
| | | .3563 | .9540 | .236 | .602 | .795 |
| | | .3591 | .9544 | .238 | .614 | .759 |
| | | .3625 | .9548 | .236 | .602 | .794 |
| | | .3653 | .9551 | .247 | .618 | .800 |
| | | .3681 | .9554 | .247 | .639 | .812 |
| | | .3709 | .9557 | .238 | .627 | .832 |
| | | .3750 | .9562 | .249 | .628 | .812 |
| | | .3771 | .9565 | .237 | .698 | .755 |
| | | .3806 | .9569 | .243 | .625 | .810 |
| | | .3834 | .9572 | .237 | .626 | .822 |
| | | .3861 | .9575 | .247 | .609 | .811 |
| 5- 6. | X.78 | 2443787.3711 | .8948 | .043 | .410 | .616 |
| | | .3738 | .8951 | .058 | .425 | .603 |
| | | .3773 | .8955 | .054 | .445 | .588 |
| | | .3801 | .8959 | .053 | .433 | .587 |
| | | .3829 | .8962 | .035 | .441 | .582 |
| | | .3856 | .8965 | .045 | .416 | .635 |
| | | .3884 | .8968 | .065 | .427 | .545 |
| | | .3912 | .8972 | .047 | .421 | .544 |
| | | .3940 | .8975 | .033 | .413 | .590 |
| | | .3968 | .8978 | .054 | .443 | .581 |
| 8- 9. | X.78 | 2443790.2379 | .2313 | -.116 | .272 | .434 |
| | | .2407 | .2316 | -.119 | .287 | .410 |
| | | .2442 | .2321 | -.128 | .283 | .414 |
| | | .2476 | .2325 | -.113 | .292 | .404 |
| | | .2511 | .2329 | -.122 | .272 | .412 |
| | | .2539 | .2332 | -.116 | .267 | .411 |
| | | .2567 | .2335 | -.120 | .273 | .444 |
| | | .2602 | .2339 | -.142 | .255 | .414 |
| | | .2629 | .2342 | -.114 | .271 | .431 |
| | | .2657 | .2346 | -.113 | .277 | .439 |
| 9-10. | X.78 | 2443791.3456 | .3613 | -.009 | .378 | .536 |
| | | .3483 | .3617 | -.009 | .380 | .522 |
| | | .3518 | .3621 | -.014 | .384 | .517 |
| | | .3546 | .3624 | -.010 | .325 | .551 |
| | | .3574 | .3627 | -.007 | .386 | .520 |
| | | .3733 | .3646 | -.027 | .379 | .493 |
| | | .3761 | .3649 | .002 | .387 | .546 |
| | | .3789 | .3652 | -.001 | .387 | .531 |
| | | .3824 | .3657 | -.007 | .384 | .537 |
| 10-11. | X.78 | 2443792.4463 | .4905 | .209 | .596 | .744 |
| | | .4497 | .4909 | .218 | .590 | .776 |
| | | .4525 | .4913 | .206 | .583 | .740 |
| | | .4553 | .4916 | .207 | .592 | .725 |
| | | .4581 | .4919 | .216 | .577 | .724 |
| | | .4609 | .4922 | .215 | .584 | .739 |
| | | .4636 | .4926 | .209 | .591 | .737 |
| | | .4657 | .4928 | .212 | .580 | .707 |
| | | .4685 | .4931 | .207 | .585 | .701 |
| | | .4713 | .4935 | .201 | .601 | .655 |
| 25-26. | XI.78 | 2443838.2641 | .8687 | -.011 | .374 | .567 |
| | | .2673 | .8691 | -.003 | .377 | .542 |
| | | .2714 | .8695 | -.008 | .367 | .541 |
| | | .2748 | .8699 | -.020 | .365 | .570 |
| | | .2769 | .8702 | -.009 | .387 | .538 |
| | | .2797 | .8705 | -.004 | .382 | .519 |
| | | .2825 | .8708 | .000 | .366 | .511 |
| 26-27. | XI.78 | 2443839.2401 | .9832 | .249 | .630 | .814 |
| | | .2429 | .9836 | .252 | .645 | .782 |
| | | .2457 | .9839 | .231 | .725 | .877 |
| | | .2484 | .9842 | .197 | .590 | .816 |
| | | .2519 | .9846 | .096 | .591 | .840 |
| | | .2554 | .9850 | .240 | .773 | .823 |
| | | .2575 | .9853 | .217 | .616 | .830 |
| | | .2603 | .9856 | .235 | .646 | .773 |
| | | .2630 | .9859 | .256 | .649 | .809 |
| | | .2658 | .9862 | .244 | .638 | .825 |
| 25-26. | I.79 | 2443899.2376 | .0258 | .283 | .627 | .779 |
| | | .2404 | .0261 | .256 | .655 | .982 |
| | | .2445 | .0266 | .275 | .635 | .781 |
| | | .2473 | .0269 | .270 | .601 | .814 |
| | | .2501 | .0272 | .250 | .618 | .778 |
| | | .2529 | .0276 | .263 | .623 | .763 |
| | | .2564 | .0280 | .247 | .607 | .766 |
| | | .2689 | .0294 | .057 | .416 | -.156 |
| 26-27. | I.79 | 2443900.2014 | .1389 | -.021 | .358 | .499 |
| | | .2042 | .1392 | -.021 | .381 | .496 |
| | | .2069 | .1395 | -.010 | .380 | .521 |
| | | .2097 | .1399 | -.025 | .361 | .504 |
| | | .2118 | .1401 | -.030 | .365 | .497 |
| | | .2146 | .1405 | -.023 | .352 | .495 |
| | | .2174 | .1408 | -.013 | .399 | .470 |
| | | .2194 | .1410 | -.027 | .374 | .517 |
| | | .2222 | .1413 | -.008 | .364 | .507 |
| | | .2250 | .1417 | -.013 | .379 | .492 |
| 29-30. | I.79 | 2443903.2124 | .4923 | .188 | .589 | .739 |
| | | .2152 | .4927 | .203 | .596 | .744 |
| | | .2180 | .4930 | .205 | .608 | .730 |
| | | .2207 | .4933 | .185 | .591 | .718 |
| | | .2235 | .4936 | .222 | .617 | .713 |
| | | .2263 | .4940 | .218 | .598 | .733 |
| | | .2291 | .4943 | .210 | .576 | .724 |
| | | .2318 | .4946 | .199 | .603 | .720 |
| | | .2360 | .4951 | .203 | .596 | .721 |
| | | .2388 | .4954 | .165 | .596 | .685 |
| 28-29. | X.79 | 2444175.3866 | .4403 | .147 | .534 | .722 |
| | | .3894 | .4406 | .151 | .539 | .715 |
| | | .3922 | .4410 | .142 | .548 | .623 |
| | | .3950 | .4413 | .144 | .564 | .625 |
| | | .3977 | .4416 | .160 | .574 | .659 |
| | | .3998 | .4419 | .164 | .563 | .646 |
| | | .4026 | .4422 | .162 | .554 | .680 |
| 17-18. | XI.79 | 2444195.2726 | .7746 | -.111 | .299 | .435 |
| | | .2761 | .7750 | -.097 | .309 | .454 |
| | | .2789 | .7752 | -.093 | .302 | .469 |
| | | .2817 | .7756 | -.088 | .307 | .481 |
| | | .2845 | .7760 | -.087 | .335 | .465 |
| | | .2872 | .7763 | -.087 | .316 | .453 |
| | | .2907 | .7767 | -.082 | .299 | .406 |
| | | .2935 | .7770 | -.081 | .306 | .414 |
| | | .2963 | .7773 | -.091 | .321 | .496 |
| | | .2990 | .7777 | -.092 | .294 | .449 |
| 19-20. | XI.79 | 2444197.3212 | .0150 | .317 | .706 | .879 |
| | | .3239 | .0153 | .324 | .701 | .860 |
| | | .3267 | .0157 | .308 | .721 | .917 |
| | | .3288 | .0159 | .313 | .714 | .895 |
| | | .3316 | .0162 | .327 | .694 | .788 |
| | | .3357 | .0167 | .323 | .703 | .817 |
| 25-26. | XI.79 | 2444203.5440 | .7455 | -.102 | .307 | .419 |
| | | .5468 | .7458 | -.099 | .325 | .437 |
| | | .5495 | .7461 | -.114 | .321 | .433 |
| | | .5523 | .7464 | -.131 | .318 | .422 |
| | | .5551 | .7468 | -.102 | .338 | .412 |
| | | .5586 | .7472 | -.113 | .314 | .450 |
| | | .5627 | .7476 | -.113 | .303 | .439 |

| | | | | | |
|---|---|---|---|---|---|
| | .5655 | .7480 | -.115 | .327 | .392 |
| | .5683 | .7483 | -.112 | .318 | .397 |
| | .5711 | .7486 | -.109 | .305 | .367 |
| 26-27. XI.79 | 2444204.3724 | .8427 | -.040 | .377 | .483 |
| | .3752 | .8430 | -.033 | .386 | .484 |
| | .3787 | .8434 | -.002 | .384 | .516 |
| | .3815 | .8438 | .004 | .381 | .488 |
| | .3843 | .8441 | -.012 | .377 | .473 |
| | .3870 | .8444 | -.020 | .391 | .494 |
| | .3898 | .8447 | .002 | .379 | .509 |
| | .3926 | .8451 | .000 | .390 | .497 |
| | .3954 | .8454 | -.025 | .364 | .517 |
| | .3981 | .8457 | -.018 | .383 | .526 |
| 10-11. XII.79 | 2444218.2609 | .4729 | .203 | .599 | .706 |
| | .2637 | .4733 | .199 | .591 | .737 |
| | .2665 | .4736 | .198 | .590 | .758 |
| | .2698 | .4739 | .192 | .601 | .738 |
| | .2720 | .4742 | .205 | .592 | .718 |
| | .2748 | .4746 | .206 | .578 | .741 |
| | .2776 | .4749 | .205 | .576 | .732 |
| | .2804 | .4752 | .204 | .586 | .731 |
| | .2832 | .4756 | .214 | .595 | .748 |
| | .2859 | .4759 | .201 | .607 | .748 |
| 12-13. XII.79 | 2444220.2295 | .7040 | -.067 | .317 | .451 |
| | .2323 | .7043 | -.072 | .310 | .461 |
| | .2350 | .7047 | -.071 | .327 | .461 |
| | .2371 | .7049 | -.075 | .301 | .445 |
| | .2406 | .7053 | -.087 | .326 | .470 |
| | .2434 | .7056 | -.075 | .328 | .436 |
| | .2462 | .7060 | -.071 | .307 | .447 |
| 16-17. VII.80 | 2444437.4436 | .2007 | -.122 | .267 | .390 |
| | .4463 | .2011 | -.144 | .243 | .385 |
| | .4491 | .2014 | -.119 | .250 | .389 |
| | .4519 | .2017 | -.137 | .248 | .380 |
| | .4547 | .2020 | -.131 | .352 | .390 |
| | .4575 | .2024 | -.132 | .251 | .382 |
| | .4602 | .2027 | -.134 | .282 | .350 |
| 17-18. VII.80 | 2444438.4419 | .3179 | -.061 | .316 | .460 |
| | .4446 | .3182 | -.062 | .345 | .457 |
| | .4474 | .3186 | -.064 | .320 | .458 |
| | .4509 | .3190 | -.083 | .332 | .450 |
| | .4537 | .3193 | -.058 | .335 | .462 |
| | .4565 | .3196 | -.048 | .372 | .506 |
| | .4592 | .3200 | -.039 | .398 | .456 |
| | .4620 | .3203 | -.055 | .314 | .473 |
| | .4648 | .3206 | -.050 | .323 | .484 |
| 21-22. VII.80 | 2444442.4593 | .7895 | -.101 | .279 | .450 |
| | .4621 | .7898 | -.115 | .303 | .388 |
| | .4649 | .7901 | -.099 | .242 | .428 |
| | .4677 | .7905 | -.125 | .271 | .438 |
| | .4704 | .7908 | -.113 | .297 | .407 |
| | .4732 | .7911 | -.114 | .306 | .390 |
| | .4760 | .7915 | -.096 | .360 | .432 |
| 7- 8.VIII.80 | 2444459.4276 | .7812 | -.044 | .328 | .471 |
| | .4304 | .7816 | -.034 | .326 | .456 |
| | .4331 | .7819 | -.054 | .315 | .463 |
| | .4359 | .7822 | -.042 | .317 | .469 |
| | .4387 | .7825 | -.054 | .338 | .459 |

| | | | | | | |
|---|---|---|---|---|---|---|
| | .4415 | .7829 | -.055 | .328 | .460 |
| | .4443 | .7832 | -.055 | .339 | .479 |
| | .4470 | .7835 | -.051 | .346 | .482 |
| | .4498 | .7838 | -.057 | .324 | .468 |
| | .4526 | .7842 | -.074 | .313 | .457 |
| 15-16.VIII.80 | 2444467.4016 | .7172 | -.077 | .314 | .438 |
| | .4044 | .7176 | -.082 | .307 | .447 |
| | .4072 | .7179 | -.082 | .312 | .447 |
| | .4106 | .7183 | -.078 | .308 | .437 |
| | .4134 | .7186 | -.076 | .302 | .445 |
| | .4162 | .7189 | -.073 | .314 | .448 |
| | .4190 | .7193 | -.081 | .327 | .439 |
| | .4217 | .7196 | -.079 | .329 | .449 |
| | .4245 | .7199 | -.080 | .328 | .437 |
| | .4273 | .7203 | -.083 | .308 | .423 |
| 17-18.VIII.80 | 2444469.3343 | .9441 | .211 | .585 | .758 |
| | .3371 | .9444 | .233 | .620 | .785 |
| | .3404 | .9448 | .220 | .612 | .751 |
| | .3468 | .9456 | .237 | .633 | .795 |
| 18-19.VIII.80 | 2444470.4310 | .0728 | .127 | .509 | .680 |
| | .4337 | .0731 | .142 | .512 | .663 |
| | .4372 | .0736 | .140 | .499 | .670 |
| | .4400 | .0739 | .142 | .513 | .671 |
| | .4428 | .0742 | .146 | .521 | .679 |
| | .4455 | .0745 | .130 | .518 | .680 |
| | .4490 | .0749 | .139 | .513 | .634 |
| | .4518 | .0753 | .119 | .503 | .651 |
| | .4544 | .0756 | .120 | .503 | .672 |
| | .4567 | .0758 | .124 | .505 | .670 |
| 28-29. XI.80 | 2444572.2633 | .0260 | .246 | .651 | .850 |
| | .2786 | .0278 | .257 | .663 | .814 |
| | .2814 | .0281 | .245 | .643 | .808 |
| | .2842 | .0284 | .252 | .651 | .823 |
| | .2876 | .0288 | .232 | .635 | .804 |
| | .2904 | .0292 | .238 | .626 | .798 |
| | .2932 | .0295 | .235 | .625 | .802 |
| | .2953 | .0297 | .240 | .632 | .797 |
| | .2980 | .0300 | .245 | .636 | .802 |
| | .3008 | .0304 | .240 | .636 | .789 |
| 26-27. XII.80 | 2444600.1990 | .3051 | -.104 | .281 | .459 |
| | .2025 | .3055 | -.091 | .310 | .457 |
| | .2053 | .3058 | -.074 | .320 | .425 |
| | .2081 | .3062 | -.092 | .312 | .449 |
| | .2108 | .3064 | -.109 | .313 | .444 |
| | .2143 | .3068 | -.087 | .298 | .445 |
| | .2171 | .3072 | -.091 | .303 | .441 |
| 6- 7. II.81 | 2444642.2203 | .2376 | -.138 | .247 | .359 |
| | .2237 | .2380 | -.133 | .250 | .355 |
| | .2272 | .2384 | -.130 | .259 | .343 |
| | .2300 | .2387 | -.126 | .266 | .329 |
| | .2334 | .2391 | -.128 | .252 | .335 |
| | .2362 | .2394 | -.138 | .246 | .355 |
| | .2404 | .2399 | -.144 | .246 | .350 |
| | .2425 | .2402 | -.152 | .247 | .351 |
| | .2453 | .2405 | -.133 | .247 | .343 |
| | .2480 | .2408 | -.131 | .237 | .345 |
| 22-23. IX.81 | 2444870.2894 | .0085 | .299 | .648 | .838 |

| | | | | | | |
|---|---|---|---|---|---|---|
| | | .2922 | .0088 | .284 | .664 | .858 |
| | | .2957 | .0092 | .298 | .677 | .841 |
| | | .2984 | .0095 | .284 | .652 | .863 |
| | | .3012 | .0099 | .288 | .664 | .840 |
| | | .3040 | .0102 | .297 | .685 | .859 |
| | | .3061 | .0104 | .306 | .667 | .845 |
| | | .3089 | .0108 | .290 | .678 | .833 |
| | | .3116 | .0111 | .283 | .654 | .836 |
| 23-24. | IX.81 | 2444871.2999 | .1271 | .006 | .389 | .536 |
| | | .3027 | .1274 | .006 | .387 | .559 |
| | | .3055 | .1277 | .020 | .377 | .387 |
| | | .3090 | .1282 | .009 | .387 | .364 |
| | | .3117 | .1285 | .014 | .413 | .398 |
| | | .3145 | .1288 | .010 | .386 | .393 |
| 24-25. | IX.81 | 2444872.3159 | .2463 | -.111 | .259 | .398 |
| | | .3187 | .2467 | -.101 | .284 | .423 |
| | | .3221 | .2471 | -.107 | .275 | .427 |
| | | .3256 | .2475 | -.115 | .288 | .437 |
| | | .3305 | .2481 | -.119 | .278 | .375 |
| | | .3333 | .2484 | -.117 | .277 | .376 |
| 25-26. | XI.81 | 2444934.1987 | .5102 | .154 | .559 | .712 |
| | | .2015 | .5105 | .195 | .587 | .725 |
| | | .2050 | .5109 | .209 | .585 | .737 |
| | | .2085 | .5113 | .199 | .599 | .762 |
| | | .2112 | .5117 | .213 | .605 | .769 |
| | | .2140 | .5120 | .203 | .610 | .746 |
| | | .2168 | .5123 | .205 | .592 | .725 |
| | | .2196 | .5126 | .210 | .593 | .732 |
| | | .2223 | .5130 | .210 | .588 | .733 |
| | | .2251 | .5133 | .218 | .577 | .752 |
| 26-27. | XI.81 | 2444935.1550 | .6224 | -.042 | .339 | .509 |
| | | .1575 | .6227 | -.049 | .328 | .496 |
| | | .1605 | .6231 | -.052 | .335 | .462 |
| | | .1633 | .6234 | -.042 | .344 | .490 |
| | | .1661 | .6237 | -.053 | .339 | .471 |
| | | .1689 | .6242 | -.051 | .340 | .481 |
| | | .1717 | .6244 | -.038 | .340 | .491 |
| 24-25. | I.82 | 2444994.2175 | .5552 | .110 | .366 | .289 |
| | | .2202 | .5556 | .140 | .361 | .289 |
| | | .2237 | .5560 | .110 | .377 | .255 |
| | | .2272 | .5564 | .103 | .385 | .262 |
| | | .2307 | .5568 | .095 | .369 | .298 |
| | | .2334 | .5571 | .103 | .377 | .285 |
| | | .2362 | .5574 | .102 | .380 | .293 |
| | | .2390 | .5578 | .108 | .374 | .323 |
| | | .2430 | .5582 | .099 | .401 | .347 |
| | | .2466 | .5586 | .111 | .401 | .300 |
| | | .2494 | .5589 | .104 | .408 | .272 |
| | | .2522 | .5593 | .122 | .403 | .307 |
| | | .2557 | .5597 | .092 | .412 | .325 |
| | | .2584 | .5600 | .085 | .427 | .289 |
| | | .2612 | .5604 | .102 | .411 | .329 |
| | | .2709 | .5615 | .114 | .418 | .314 |
| 18-19. | IX.82 | 2445231.3414 | .3890 | .010 | .392 | .546 |
| | | .3442 | .3893 | .015 | .415 | .572 |
| | | .3469 | .3897 | .018 | .421 | .558 |
| | | .3497 | .3900 | -.036 | .358 | .544 |
| | | .3532 | .3904 | .015 | .390 | .529 |
| | | .3560 | .3907 | .008 | .384 | .548 |
| | | .3601 | .3912 | .002 | .398 | .545 |
| 15-16. | X.82 | 2445258.2845 | .5517 | .121 | .495 | .659 |
| | | .2872 | .5519 | .123 | .514 | .665 |
| | | .2907 | .5519 | .129 | .512 | .683 |
| | | .2928 | .5526 | .130 | .518 | .674 |
| | | .2949 | .5528 | .130 | .502 | .661 |
| | | .2983 | .5532 | .124 | .508 | .657 |
| | | .3011 | .5535 | .119 | .554 | .617 |
| | | .3032 | .5538 | .127 | .516 | .625 |
| 16-17. | X.82 | 2445259.2699 | .6673 | -.055 | .321 | .466 |
| | | .2733 | .6676 | -.056 | .314 | .477 |
| | | .2782 | .6682 | -.054 | .322 | .477 |
| | | .2810 | .6686 | -.048 | .330 | .494 |
| | | .2838 | .6689 | -.066 | .323 | .432 |
| | | .2872 | .6693 | -.058 | .325 | .442 |
| | | .2900 | .6696 | -.056 | .324 | .454 |
| | | .2921 | .6699 | -.047 | .341 | .468 |
| | | .2949 | .6702 | -.060 | .318 | .434 |
| 21-22. | X.82 | 2445264.2943 | .2570 | -.132 | .240 | .382 |
| | | .2971 | .2574 | -.142 | .240 | .411 |
| | | .2998 | .2577 | -.132 | .249 | .382 |
| | | .3026 | .2581 | -.136 | .244 | .365 |
| | | .3061 | .2584 | -.138 | .250 | .387 |
| | | .3082 | .2587 | -.127 | .252 | .384 |
| | | .3109 | .2590 | -.135 | .254 | .407 |
| 24-25. | X.82 | 2445267.2998 | .6099 | .005 | .157 | .272 |
| | | .3026 | .6102 | .018 | .373 | .501 |
| | | .3054 | .6104 | .016 | .382 | .541 |
| | | .3075 | .6108 | .020 | .379 | .504 |
| | | .3102 | .6110 | .017 | .355 | .500 |
| | | .3130 | .6114 | .012 | .356 | .547 |
| | | .3158 | .6117 | .008 | .382 | .535 |
| | | .3179 | .6120 | .009 | .388 | .515 |
| | | .3214 | .6123 | -.003 | .371 | .513 |
| | | .3241 | .6127 | .001 | .381 | .525 |
| 25-26. | X.82 | 2445268.3513 | .7332 | -.109 | .265 | .416 |
| | | .3541 | .7336 | -.100 | .284 | .448 |
| | | .3569 | .7339 | -.104 | .271 | .393 |
| | | .3597 | .7343 | -.106 | .283 | .386 |
| | | .3624 | .7345 | -.106 | .296 | .396 |
| | | .3680 | .7352 | -.106 | .277 | .404 |
| | | .3701 | .7355 | -.110 | .295 | .397 |
| 26-27. | X.82 | 2445269.3972 | .8560 | -.060 | .346 | .504 |
| | | .3999 | .8564 | -.051 | .363 | .481 |
| | | .4034 | .8567 | -.043 | .358 | .481 |
| | | .4062 | .8571 | -.045 | .387 | .482 |
| | | .4090 | .8574 | -.036 | .349 | .503 |
| | | .4117 | .8578 | -.034 | .349 | .485 |
| | | .4138 | .8580 | -.035 | .361 | .472 |
| 7- 8. | XII.82 | 2445311.2173 | .7649 | -.134 | .264 | .410 |
| | | .2201 | .7652 | -.124 | .252 | .406 |
| | | .2249 | .7658 | -.134 | .285 | .315 |
| | | .2284 | .7662 | -.125 | .262 | .293 |
| | | .2312 | .7665 | -.117 | .265 | .404 |
| | | .2358 | .7670 | -.115 | .251 | .373 |
| | | .2381 | .7673 | -.118 | .265 | .413 |

| | | | | | |
|---|---|---|---|---|---|
| 14-15. XII.82 | 2445318.1871 | .5830 | .028 | .413 | .569 |
| | .1899 | .5832 | .039 | .412 | .541 |
| | .1927 | .5837 | .036 | .382 | .516 |
| | .1948 | .5839 | .037 | .377 | .588 |
| | .1975 | .5843 | .043 | .413 | .533 |
| | .2003 | .5845 | .023 | .406 | .543 |
| | .2031 | .5849 | .021 | .388 | .537 |
| | .2059 | .5852 | .049 | .401 | .544 |
| | .2080 | .5855 | .026 | .398 | .553 |
| | .2107 | .5858 | .016 | .421 | .592 |
| | .2135 | .5862 | .035 | .402 | .536 |
| | .2156 | .5864 | .036 | .393 | .560 |
| | .2218 | .5871 | -.012 | .340 | .526 |
| | .2246 | .5875 | .017 | .418 | .557 |
| 17-18. XII.82 | 2445321.2001 | .9367 | .129 | .503 | .650 |
| | .2029 | .9370 | .137 | .548 | .644 |
| | .2071 | .9375 | .132 | .501 | .648 |
| | .2105 | .9380 | .118 | .492 | .653 |
| | .2140 | .9383 | .110 | .509 | .646 |
| | .2175 | .9388 | .112 | .518 | .676 |
| | .2237 | .9395 | .132 | .491 | .650 |
| | .2265 | .9398 | .127 | .511 | .665 |
| | .2286 | .9401 | .146 | .524 | .672 |
| | .2314 | .9403 | .124 | .510 | .678 |
| 13-14. I.83 | 2445348.2259 | .1090 | -.005 | .408 | .515 |
| | .2287 | .1094 | .002 | .416 | .512 |
| | .2315 | .1097 | .008 | .403 | .543 |
| | .2342 | .1099 | .002 | .397 | .534 |
| | .2370 | .1103 | .011 | .400 | .503 |
| | .2419 | .1109 | .011 | .395 | .519 |
| | .2474 | .1115 | -.003 | .396 | .502 |
| 3- 4. II.83 | 2445369.1721 | .5677 | .057 | .431 | .533 |
| | .1749 | .5680 | .062 | .446 | .566 |
| | .1784 | .5684 | .070 | .467 | .577 |
| | .1812 | .5687 | .059 | .455 | .570 |
| | .1839 | .5691 | .056 | .435 | .568 |
| | .1867 | .5694 | .074 | .431 | .564 |
| | .1895 | .5698 | .052 | .412 | .527 |
| | .1916 | .5700 | .065 | .420 | .505 |
| | .1943 | .5702 | .050 | .430 | .553 |
| | .1971 | .5706 | .053 | .431 | .553 |
| 4- 5. II.83 | 2445370.1860 | .6867 | -.062 | .296 | .440 |
| | .1888 | .6870 | -.046 | .317 | .434 |
| | .1916 | .6874 | -.081 | .320 | .420 |
| | .1943 | .6876 | -.069 | .304 | .423 |
| | .1971 | .6880 | -.064 | .307 | .434 |
| | .1992 | .6882 | -.044 | .326 | .434 |
| | .2027 | .6887 | -.041 | .325 | .433 |
| | .2055 | .6890 | -.084 | .301 | .426 |
| | .2082 | .6893 | -.091 | .329 | .418 |
| | .2103 | .6895 | -.079 | .315 | .449 |
| 13-14. IX.83 | 2445591.4030 | .6533 | -.053 | .327 | .508 |
| | .4072 | .6538 | -.057 | .317 | .484 |
| | .4106 | .6542 | -.046 | .310 | .485 |
| | .4141 | .6546 | -.044 | .339 | .508 |
| | .4169 | .6549 | -.056 | .358 | .465 |
| | .4211 | .6554 | -.078 | .344 | .488 |
| | .4245 | .6559 | -.069 | .330 | .492 |
| 14-15. IX.83 | 2445592.4010 | .7704 | -.105 | .257 | .413 |
| | .4038 | .7708 | -.091 | .280 | .413 |
| | .4073 | .7711 | -.080 | .293 | .404 |
| | .4107 | .7716 | -.075 | .299 | .401 |
| | .4163 | .7722 | -.088 | .267 | .419 |
| | .4205 | .7728 | -.091 | .288 | .435 |
| | .4239 | .7731 | -.077 | .300 | .434 |
| 16-17. IX.83 | 2445594.4643 | .0126 | .281 | .672 | .880 |
| | .4678 | .0131 | .263 | .671 | .815 |
| | .4706 | .0134 | .256 | .685 | .818 |
| | .4726 | .0136 | .271 | .689 | .873 |
| | .4754 | .0139 | .273 | .663 | .824 |
| | .4782 | .0142 | .253 | .658 | .832 |
| | .4810 | .0146 | .264 | .640 | .851 |
| | .4844 | .0149 | .258 | .637 | .864 |
| 30-31. I.84 | 2445730.1769 | .9427 | .200 | .606 | .756 |
| | .1797 | .9430 | .195 | .579 | .743 |
| | .1824 | .9433 | .203 | .576 | .740 |
| | .1852 | .9436 | .203 | .603 | .683 |
| | .1894 | .9441 | .251 | .636 | .739 |
| | .1936 | .9447 | .191 | .592 | .735 |
| | .1984 | .9451 | .198 | .603 | .767 |
| | .2019 | .9456 | .214 | .605 | .714 |
| | .2047 | .9460 | .184 | .606 | .712 |
| | .2074 | .9462 | .208 | .611 | .763 |
| | .2158 | .9472 | .206 | .547 | .690 |
| | .2186 | .9476 | .222 | .575 | .740 |
| | .2213 | .9478 | .212 | .593 | .739 |
| | .2241 | .9482 | .216 | .635 | .783 |
| | .2276 | .9487 | .225 | .597 | .740 |
| | .2304 | .9489 | .225 | .612 | .750 |
| | .2324 | .9491 | .221 | .611 | .777 |
| | .2352 | .9495 | .215 | .628 | .743 |
| | .2387 | .9449 | .232 | .602 | .740 |
| | .2415 | .9503 | .241 | .636 | .740 |
| | .2442 | .9505 | .233 | .614 | .754 |
| 20-21. XII.84 | 2446055.2805 | .1036 | .001 | .397 | .483 |
| | .2832 | .1039 | .024 | .429 | .529 |
| | .2864 | .1043 | .003 | .391 | .494 |
| | .2889 | .1046 | .005 | .366 | .508 |
| | .2994 | .1058 | .005 | .399 | .549 |
| | .3025 | .1062 | .002 | .411 | .459 |
| | .3051 | .1065 | .010 | .386 | .476 |